\documentclass{article}
\usepackage[utf8]{inputenc}
\usepackage{amsmath}
\usepackage{graphicx}
\usepackage{caption, subcaption}
\usepackage{float}
\usepackage{indentfirst}
\usepackage{booktabs}

\usepackage{multirow}
\usepackage{authblk}
\usepackage{tabularx}
\usepackage{geometry}
\usepackage{appendix}
\pdfoutput=1 

\title{Simulation and application of COVID-19 compartmental model using Physics-informed Neural Network}
\author[1]{Jinhuan Ke}
\author[2]{Jiahao Ma}
\author[3]{Xiyu Yin}
\author[4]{Robin Singh}
\affil[1]{Department of Statistics, University of Michigan, MI, USA, 48105}
\affil[2]{Department of Marine Science and Technology, Harbin Institute of Technology University, Weihai, China, 264209}
\affil[3]{Department of Biological Science, University of Connecticut, CT, USA, 06269}
\affil[4]{Department of Mechanical Engineering, Massachusetts Institute of Technology, MA, USA, 02139}
\date{}

\begin{document}

\maketitle

\begin{abstract}
COVID-19 pandemic has had a disruptive and irreversible impact globally, yet traditional epidemiological modeling approaches such as the susceptible-infected-recovered (SIR) model have exhibited limited effectiveness in forecasting of the up-to-date pandemic situation. In this work, susceptible-vaccinated-exposed-infected-dead-recovered (SVEIDR) model and its variants -- aged and vaccination-structured SVEIDR models -- are introduced to encode the effect of social contact for different age groups and vaccination status. Then, we implement the physics-informed neural network (PiNN) on both simulated and real-world data. The PiNN model enables robust analysis of the dynamic spread, prediction, and parameter optimization of the COVID-19 compartmental models. The models exhibit relative root mean square error (RRMSE) of $<4\%$ for all components and provide incubation, death, and recovery rates of $\gamma= 0.0130$, $\lambda=0.0001$, and $\rho=0.0037$, respectively, for the first 310 days of the epidemic in the US with RRMSE of $<0.35\%$ for all components. To further improve the model performance, temporally varying parameters can be included, such as vaccination, transmission, and incubation rates. Our implementation highlights PiNN as a reliable candidate approach for forecasting real-world data and can be applied to other compartmental model variants of interest.
\end{abstract}

\section{Introduction}

COVID-19 is an infectious epidemic caused by a new Coronavirus virus(SARS-Cov-2). Since the outbreak of COVID-19 at the end of 2019, it has spread throughout the world. As of May 20, 2022, there are more than 5 hundred million 
confirmed cases and over 6 million 
deaths worldwide. COVID- 19 has had a significant impact on the global economy, human health, and daily life. In response to this pandemic crisis, researchers and policymakers worldwide have been working hard to study and develop countermeasures and solutions against COVID-19 to control the epidemic and reduce its impact on human health and the economy.

There have been many studies on COVID-19 by modeling the spread of the virus and predicting the number of people infected by the virus. Since the mathematical theory of epidemiology was proposed by Kermack in 1927 \cite{kermack1927contribution}
, the mathematical model of infectious diseases has been used as an essential tool for epidemiological feature analysis and transmission analysis. The most commonly used models are SIR (Susceptible, Infected, and Recovered) and SEIR (Susceptible, Exposed, Infected, and Recovered) models, which belong to compartmental models based on the set of mathematical differential equations. Most studies are based on these two models with corresponding modifications and adjustments. Liao et al. \cite{liao2020tw} proposed a generalized adaptive SIR prediction model based on a time window that introduces a time window mechanism for dynamic data analysis, which can dynamically measure the number of initial infections and the exponential growth rate. Pushpendra et al. \cite{singh2021generalized} proposed a generalized SIR (GSIR) model including multi-day reported cases. Experimental evaluations of COVID-19 outbreak data in different countries show that the model is capable of continuous prediction and monitoring of COVID-19 outbreaks. Cooper et al. \cite{cooper2020sir} added the number of susceptible people as a variable on the basis of the SIR model, providing a theoretical framework for studying the spread of COVID-19 in the community. Hakimeh et al. \cite{mohammadi2021fractional} established a fractional-order SIRD (susceptibility, infection, and death) mathematical model of COVID-19 transmission using Caputo derivatives. The second wave of outbreaks in Iran and Japan was predicted through numerical simulations of different order derivatives. Ramezani et al. \cite{ramezani2021novel} proposed a variant of the SEIRD (susceptibility, infection, recovery, and death) model, which captures the nonlinear behavior of the COVID-19 pandemic while accounting for asymptomatic infected individuals. For modifying the compartmental model embed with social connection, \cite{ram2021modified} makes progress on separating people to be different age groups and built a modified age-structured SIR model. These paper fit and predict data well yet are lack of consideration on vaccination's contribution in the COVID-19 infectious situation, and also require more analysis on modeling contact and interaction among diverse people.

As an emerging method, deep learning technology has been widely used in the analysis and forecast of the COVID-19 epidemic. Vinay et al. \cite{CHIMMULA2020109864} first used LSTM (Long Short Term Memory) deep learning model, to model the spread of infectious diseases in Canada for the prediction The severity of COVID-19. Khondoker et al. \cite{NABI2021104137} studied a group of four deep learning models: LSTM,  MCNN(Multivariate convolutional neural network), GRU(Gated recurrent unit), and CNN(Convolutional neural network), the results show that CNN outperforms other deep learning models in terms of validation accuracy and prediction consistency. Smail et al. \cite{KIRBAS2020110015} did some study about another group of models: NARNN(Nonlinear autoregressive neural network), ARIMA(Autoregressive integrated moving average method), and LSTM methods to model COVID-19 confirmed, and the result came out that LSTM is the most accurate model among those four. Jayanthi et al. \cite{DEVARAJ2021103817} used LSTM, Stacked LSTM, ARIMA, and prophet models to analyze and predict the cumulative global confirmed cases, deaths and recoveries. The results show that the Stacked LSTM algorithm achieves higher accuracy with less than 2 percent error compared to other considered methods. As can be seen from the above studies, deep learning can provide reliable single-day forecasting results. Nevertheless, there are two main problems with these methods. The first is that these purely numerical fitting methods do not properly capture popular trends in the propagation process. \cite{NABI2021104137} Another is that the temporal pattern of the number of infected people is very simple. It is difficult for these methods to find the correct pattern of long-term changes, resulting in effective predictions only in the short term. A model that provides long-term forecasts is clearly essential if decision-making and planning is to be more effective.

Mathematical models of infectious diseases can be used to predict the spread of epidemics, but the introduction of model parameters is based on many assumptions. Various unknown parameters need to be estimated by model fitting, which also makes the model more uncertain \cite{holmdahl2020wrong}. In addition, due to the long duration of COVID-19, many different factors are affecting the parameters of the model, therefore it is difficult to use only a single model to adapt well to the actual situation \cite{panovska2020can}. Then, whether the numerical simulation after the preset conditions (including parameters, initial values of each compartment and channels between compartments) or the modeling and prediction based on real data need to be realized by calculation tools.

Recently, Neural network (NN) is considered to be one of the fastest growing fields, especially training method of weak supervision or unsupervised that make neural networks apply across a broad range of fields. Earlier, many researchers have proposed Physics-informed Neural Network (PiNN) \cite{dissanayake1994neural} to automatically search for approximate solution of differential equations representing physical processes. After that, with the rapid development of software (e.g. pytorch\cite{paszke2019pytorch}) and hardware (e.g. RTX3090\cite{shacklett2021large}) and the advent of the era of big data, different kinds of Informed Neural Network (INN) has received a lot of research and gradually developed into an important tool to solve many physics, statistics and engineering problems\cite{raissi2017physics}. Different from pure data-driven model or pure knowledge driven model, it can be used as a "bridge" to connect data methods with traditional knowledge driven equations. This kind of neural network is utilized and experiments, focusing on SIR model, have been carried out in papers such as \cite{DBLP:journals/corr/abs-2110-05445}, and further implementations of PiNN are desired application on more complex and informative compartmental models.

To effectively predict COVID-19, more advanced methods should be applied to consider various factors in virus transmission. We propose a COVID-19 prediction model, SVEIRD-PiNN (Susceptible, Vaccinated, Exposed, Infected, Recovered, and Dead with Physic-Informed Neural Network). In our work, two directions, including parameter-driven simulation direction studying on simulated data and data-driven direction studying on real-world data, are learned by PiNN, where SVEIDR (Susceptible, Vaccinated, Exposed, Infected, Dead, and Recovered) model (section \ref{Basic SVEIDR Model}) and its variant models are mainly focused on. To analyze the effect of COVID-19 on different age groups, we introduce a contact network in \cite{ram2021modified}, and derive the first variant:  Aged-structured SVEIDR model (section \ref{Age structured Model}) where diverse transmission rate are encoded in the differential equations for different age groups. Also, since vaccination are gradually deriving different status, such as doses, brands, boosters and so on, necessity is being aware to integrate the vaccination status to the COVID-19 effect. In this paper, a vaccination-structured SVEIDR model (section \ref{Vaccination structured Model}) is introduced to focus on this issue, whose construction is not only inspired from but also linked with the Aged-structured SVEIDR model. After model construction, the paper as well makes analysis on the mathematical diagnosis (section \ref{Diagnosis}). Then we apply PiNN (section  \ref{PINN}) to SVEIDR models and implement it on both simulated and real-world data (section \ref{Implement}).

\section{Methods}
\label{methods}

\subsection{Basic SVEIDR model}
\label{Basic SVEIDR Model}
\begin{figure}[H]
    \centering
    \includegraphics[scale=0.5]{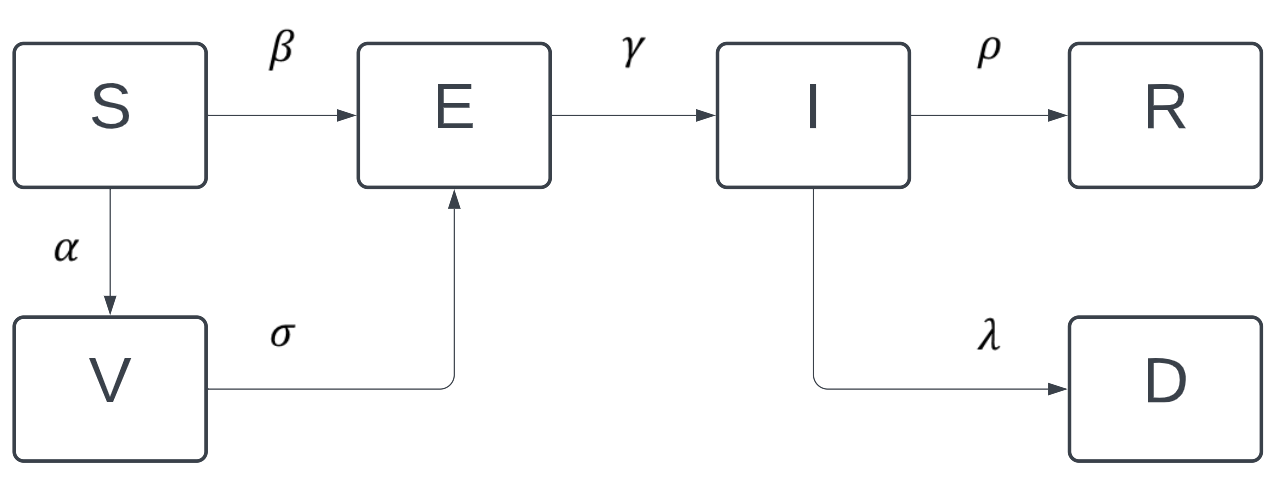}
    \caption{Basic SVEIDR model: Susceptible (S), vaccinated (V), exposed (E), infected (I), dead (D), and recovered (R) people transform amongst each other at different rates (Greek symbols).}
    \label{fig:wkfl}
\end{figure}

Population is divided into several compartments (susceptible, vaccinated, exposed, infectious, dead, recovered people) to formulate the interaction and prediction of the infectious disease analysis. Susceptible people refer to those original without being  infected or infectiously contacted, exposed people represent those who are infected but not yet infectious, and infectious compartment refers to the infectious public. Vaccinated compartment is considered to be citizens who are vaccinated and have certain immunity towards the infection. Also, people recover from or die of the disease are described as recovered and dead compartment respectively. Different compartment interacts with other and can transform from one to another with diverse rate. $S(t)$ represents the number of susceptible people at time $t$, and we also use $S$ for simplification. The detailed description of the basic SVEIDR model is shown in the following ordinary differential equations and the compartmental figure \ref{fig:wkfl}.

- $S(t)$: Nummber of the  susceptible

- $E(t)$: Nummber of the exposed

- $I(t)$: Nummber of the  infectious

- $V(t)$; Nummber of the  vaccinated

- $R(t)$: Nummber of the  recovered

- $D(t)$: Nummber of the  dead

- $N$: Number of total people

\begin{align}
\label{f_basic}
\begin{split}
     \frac{dS(t)}{dt} &= -\beta S I -\alpha S\\
     \frac{dV(t)}{dt} &= \alpha S -\sigma \beta VI\\
    \frac{dE(t)}{dt} &= \beta SI -\gamma E + \sigma \beta VI\\
    \frac{dI(t)}{dt} &= \gamma E -\lambda I -\rho I\\
    \frac{dR(t)}{dt} &= \rho I\\
    \frac{dD(t)}{dt} &= \lambda I\\
    N &= S+V+E+I+R+D
\end{split}
\end{align}
The utility of the ordinary differential equations (ODE) to describe the epidemic system can refer to \cite{martcheva2015introduction}, and the details of parameters are shown in table \ref{tab: parameter}. Note that this paper is analyzing a close epidemic system without any births and deaths that are not produced by the disease. In other words, the total number $N$ is stable.

\subsection{Age-structured SVEIDR model}
\label{Age structured Model}
In reality, the infection of COVID-19 various among different ages, taking account that people in different age groups tend to behave diverse in social life with similar aged people as well as distinct aged people. We here, referring to \cite{ram2021modified}, separate each compartment in each of the age-brackets, and utilize the age contact matrix $M$, as shown in the following. The contact matrix presents the coefficients to describe the social connection among diverse age groups (such as A groups), the higher the value of the contact coefficient $M_{ij}$ ($i,j=1:A$) is, the higher strength of interaction people tend to show in their social life, and this can make contribution to specific the infectious contact when forming the formulas.
\begin{align}
\label{contactmatrix}
    M =\begin{pmatrix}
    M_{11} & \dots& M_{1A}\\
    \vdots & & \vdots\\
    M_{A1} &\dots& M_{AA}
    \end{pmatrix}_{A \times A}
\end{align}

For each age group i, we have the following ODE (parameters described in table \ref{tab: parameter}):
\begin{align}
\label{f_aged}
\begin{split}
    \frac{dS_i(t)}{dt} &= -\beta S_i \sum_{j=1}^A M_{ij} I_j -\alpha S_i\\
    \frac{dV_i(t)}{dt} &= \alpha S_i -\sigma \beta V_i\sum_{j=1}^A M_{ij} I_j\\
    \frac{dE_i(t)}{dt} &= \beta S_i\sum_{j=1}^A M_{ij} I_j -\gamma E_i + \sigma \beta V_i\sum_{j=1}^A M_{ij} I_j\\
    \frac{dI_i(t)}{dt} &= \gamma E_i -\lambda I_i -\rho I_i\\
    \frac{dR_i(t)}{dt} &= \rho I_i\\
    \frac{dD_i(t)}{dt} &= \lambda I_i\\
    N_i &= S_i + V_i + E_i + I_i + D_i + R_i
\end{split}
\end{align}

\subsection{Vaccination-structured SVEIDR model}
The COVID-19 vaccination has brought the epidemic a new stage after the majority of citizens are vaccinated, and that's how the vaccinated compartment comes to an essential place in the SVEIDR structure model. To detailed describe how the vaccinated status affect the model, this paper develops the thought in section \ref{Age structured Model} to separate each of vaccination status, especially dose or vaccination brand, and combines them with the age groups to build a new vaccination-structured SVEIDR model.

\label{Vaccination structured Model}
For each dose or brand k, we respectively have its vaccination rate $\alpha_k$, and vaccine inefficacy $\sigma_k$:
\begin{align}
\label{f_vac}
\begin{split}
    \frac{dS_{i}(t)}{dt} &= -\beta S_{i} \sum_{j=1}^A M_{ij}  I_{j} -\sum_{k=1}^K \alpha_k S_{i}\\
     \frac{dV_{ik}(t)}{dt} &= \alpha_k S_i -\sigma_k \beta V_{ik}\sum_{j=1}^A M_{ij} I_j\\
    \frac{dE_{i}(t)}{dt} &= \beta S_{i}\sum_{j=1}^A M_{ij} I_{j} -\gamma E_{i} + \sum_{k=1}^K \sigma_k \beta V_{ik}\sum_{j=1}^A M_{ij} I_{j}\\
    \frac{dI_{i}(t)}{dt} &= \gamma E_i -\lambda I_i -\rho I_i\\
    \frac{dR_i(t)}{dt} &= \rho I_i\\
    \frac{dD_i(t)}{dt} &= \lambda I_i  \\
    N_i &= S_i + \sum_{k=1}^K V_{ik} + E_i + I_i + D_i + R_i
\end{split}
\end{align}

Apart from vaccination doses and brands, we can also assign different vaccination status as a category in K (simply one-dose, fully vaccinated, booster and so on). Further, we can extend the model to have different vaccination effectiveness that will affect the incubation rate $\gamma$, death rate $\lambda$, and recovery rate $\rho$. More details about the parameters in ODE are depicted in table \ref{tab: parameter}.

\subsection{Model mathematical analysis}
To ensure the global stability of the modified SVEIDR model we built to depict the equilibrium, the non-negativity and the basic reproduction number of the vaccinated-structured SVEIDR model (section \ref{Vaccination structured Model}) are discussed in this section.
\label{Diagnosis}
\subsubsection{Non-negativity}

Th. If $S_i(0) \geq 0, \dots, D_i(0) \geq 0$, then the solutions
of the system remain non-negative for all $t > 0$.

Proof.\\
For $S_i(t)$:
\begin{gather}
    \frac{dS_{i}(t)}{dt} = -\beta S_{i} \sum_{j=1}^A M_{ij} I_{j} -\sum_{k=1}^K \alpha_k S_{i}\\
    \rightarrow
    \frac{dS_i(t) }{S_i(t)} =  -\beta  \sum_{j=1}^A M_{ij} I_{j} -\sum_{k=1}^K \alpha_k\\
    \rightarrow
    ln(\frac{S_i(t) }{S_i(0)}) = - \int_{0}^{t} (\beta  \sum_{j=1}^A M_{ij} I_{j} +\sum_{k=1}^K \alpha_k) d\tau\\
    \rightarrow
    S_i(t) = S_i(0) exp(-\int_{0}^{t} (\beta  \sum_{j=1}^A M_{ij} I_{j} +\sum_{k=1}^K \alpha_k) d\tau)
\end{gather}
Hence, $S_i(t) \geq 0$ holds when $S_i(0) \geq 0$.

Similarly, we can obtain $E_i(0) \geq 0, \dots, D_i(0) \geq 0$.

\subsubsection{Basic reproduction number}
The basic reproduction number $R_0$ can be widely seen in the study of epidemiology, where the parameter plays the role of threshold to predict whether the infection will spread. If $R_0 > 1$, then the infected individual can averagely infect more than one person to make the epidemic situation spread, and vice versa. The following framework is from the idea given in \cite{k_filter_math9060636}, \cite{VANDENDRIESSCHE200229} and \cite{age_born_CAO2012385}.

Referring to \cite{VANDENDRIESSCHE200229}, to find the basic reproduction number $R_0$, we need to figure out the next generation matrix, whose spectral radius can be interpreted as "the typical number of secondary cases", i.e $R_0$.

For 
\begin{align}
    X = (S_1, S_2, \dots, S_A, V_{11}, \dots, V_{AK}, E_{1},\dots, E_{A}, I_1, \dots, I_A, R_1, \dots, D_A)^T
\end{align}

we have the epidemic equilibrium $X^0$,  derived by setting all the derivatives to zero with the infected compartment $I=0$. Since we are imposing a constant population $N$ and stable $N_i$ for each age group i, the Disease Free Equilibrium (DFE) can be defined as $V_{ik}^0 = N_{ik} = N_i/K$ and all other components equal to 0 for each age group i and dose group k. Note that we do not consider this to be the unique DFE of the model.
\begin{align}
    X^0 = (0, \dots, 0, N_{11}, \dots, N_{AK}, 0, \dots,0)^T
\end{align}

Here the dim of X is n=5A+AK.

Let $\hat X = (E,I)^T$, which has 2A dim. Then the vaccinated-structured SVEIDR model can be written as:

\begin{gather}
    \nabla \hat X = F(\hat X) - W(\hat X)\\
    F(\hat X)=(\beta S_{1}\sum_{j=1}^A M_{1j} I_{j}+ \sum_{k=1}^K \sigma_k \beta V_{1k}\sum_{j=1}^A M_{1j} I_{j}, \dots, \beta S_{A}\sum_{j=1}^A M_{Aj} I_{j}+ \sum_{k=1}^K \sigma_k \beta V_{Ak}\sum_{j=1}^A M_{Aj} I_{j},0, \dots, 0)^T\\
    W(\hat X)=(\gamma E_1, \dots, \gamma E_A, -\gamma E_1 + (\lambda+\rho)I_1, \dots, -\gamma E_A + (\lambda+\rho)I_A)^T
\end{gather}

Then their Jacobian matrices at the DFE are:

\begin{gather}
\begin{split}
    J_F &= \begin{pmatrix}
    0&\dots&0&
    [S_{1}^0+\sum_{k=1}^K \sigma_k V_{1k}^0]\beta M_{11}&\dots&
    [S_{1}^0+\sum_{k=1}^K \sigma_k  V_{1k}^0]\beta M_{1A}\\
    0&\dots&0&
    [S_{2}^0+\sum_{k=1}^K \sigma_k  V_{2k}^0]\beta M_{21}&\dots&
    [S_{2}^0+\sum_{k=1}^K \sigma_k  V_{2k}^0]\beta M_{2A}\\
    \vdots\\
    0&\dots&0&
    [S_{A}^0+\sum_{k=1}^K \sigma_k  V_{Ak}^0]\beta M_{A1}&\dots&
    [S_{A}^0+\sum_{k=1}^K \sigma_k \beta V_{Ak}^0]\beta M_{AA}\\
    \vdots\\
    0&\dots&0&0&\dots&0
    \end{pmatrix}\\
   & = \begin{pmatrix}
    0&\dots&0&
    \sum_{k=1}^K \sigma_k N_1 K^{-1} \beta M_{11}&\dots&
    \sum_{k=1}^K \sigma_k N_1 K^{-1}\beta M_{1A}\\
    0&\dots&0&
     \sum_{k=1}^K \sigma_k N_2 K^{-1}\beta M_{21}&\dots&
    \sum_{k=1}^K \sigma_k N_2 K^{-1}\beta M_{2A}\\
    \vdots\\
    0&\dots&0&
    \sum_{k=1}^K \sigma_k N_A K^{-1}\beta M_{A1}&\dots&
    \sum_{k=1}^K \sigma_k N_A K^{-1}\beta M_{AA}\\
    \vdots\\
    0&\dots&0&0&\dots&0
    \end{pmatrix}\\
    &=\sum_{k=1}^K \sigma_k \beta K^{-1}
    \begin{pmatrix}
    0&N \cdot M\\
    0&0
    \end{pmatrix}
\end{split}
\end{gather}
\begin{gather}
    J_W = 
    \begin{pmatrix}
    \gamma&0&\dots&\dots&\dots&\dots&0\\
    0&\gamma&0&\dots&\dots&\dots&0\\
    \vdots\\
    0&\dots&0&\gamma&0&\dots&0\\
    -\gamma&0&\dots&\dots&(\lambda+\rho)&\dots&0\\
    0&-\gamma&0&\dots&\dots&\dots&0\\
    \vdots\\
    0&\dots&0&-\gamma&0&\dots&(\lambda+\rho)
    \end{pmatrix}
\end{gather}

Here, $J_F$ and $J_W$ is $2A\times2A$ dim.

\begin{align}
\begin{split}
    J_W^{-1}&= 
    \begin{pmatrix}
    A&0\\C&D
    \end{pmatrix}^{-1}= 
    \begin{pmatrix}
    A^{-1}&0\\-D^{-1}CA^{-1}&D^{-1}
    \end{pmatrix}\\
    &=
    \begin{pmatrix}
    \gamma^{-1}&0&\dots&\dots&\dots&\dots&0\\
    0&\gamma^{-1}&0&\dots&\dots&\dots&0\\
    \vdots\\
    0&\dots&0&\gamma^{-1}&0&\dots&0\\
    (\lambda+\rho)^{-1}&0&\dots&\dots&(\lambda+\rho)^{-1}&\dots&0\\
    0&(\lambda+\rho)^{-1}&0&\dots&\dots&\dots&0\\
    \vdots\\
    0&\dots&0&(\lambda+\rho)^{-1}&0&\dots&(\lambda+\rho)^{-1}
    \end{pmatrix}
\end{split}
\end{align}

Then, according to \cite{VANDENDRIESSCHE200229}, focusing on an infected individual that introduced to a free-disease compartment $k$, the $(i,j)$ entry of $J_F$ is the rate that the infected individuals in compartment $j$ produce new infections in compartment $i$, and the $(j,k)$ entry of $J_W^{-1}$ represents the average time that the introduced individual spends in compartment $j$ during its life time. To sum up, the $(i,k)$ entry of $_F J_W^{-1}$ shows the expected number of newly infected people produced by the originally introduced infected individuals in compartment $k$, then the reproduction number $R_0$ can be calculated as the spectral radius of the next-generation matrix $J_F J_W^{-1}$.
\begin{align}
\begin{split}
    R_0 &= \Lambda \{J_F J_W^{-1}\}\\
    &= \Lambda\{ \sum_{k=1}^K \sigma_k \beta K^{-1}
    \begin{pmatrix}
    (\lambda+\rho)^{-1}N_1 M_{11} &\dots& (\lambda+\rho)^{-1}N_1  M_{1A}
    &(\lambda+\rho)^{-1}N_1  M_{11}&\dots&(\lambda+\rho)^{-1}N_1  M_{1A}\\
    \vdots\\
    (\lambda+\rho)^{-1}N_A M_{A1} &\dots& (\lambda+\rho)^{-1}N_A M_{AA}
    &(\lambda+\rho)^{-1}N_A M_{A1}&\dots&(\lambda+\rho)^{-1}N_A M_{AA}\\
    0&\dots&0&0&\dots&0\\
    \vdots\\
    0&\dots&0&0&\dots&0
    \end{pmatrix}\}\\
    &=\Lambda\{\sum_{k=1}^K \sigma_k \beta K^{-1} (\lambda+\rho)^{-1}
    \begin{pmatrix}
    N \cdot M &N \cdot M\\
    0&0
    \end{pmatrix}\}\\
    &= \Lambda\{\sum_{k=1}^K \sigma_k \beta K^{-1} (\lambda+\rho)^{-1}
    \begin{pmatrix}
    N \cdot M 
    \end{pmatrix}\}
\end{split}
\end{align}

Note that, $$N=\begin{pmatrix}
    N_1 & \dots& N_1\\
    \vdots & & \vdots\\
    N_A &\dots& N_A
    \end{pmatrix}_{A \times A}, M =\begin{pmatrix}
    M_{11} & \dots& M_{1A}\\
    \vdots & & \vdots\\
    M_{A1} &\dots& M_{AA}
    \end{pmatrix}_{A \times A}, N \cdot M = \begin{pmatrix}
    N_1 M_{11} & \dots& N_1 M_{1A}\\
    \vdots & & \vdots\\
    N_A M_{A1} &\dots& N_A M_{AA}
    \end{pmatrix}_{A \times A}$$.

And $\Lambda \{\}$ denotes the spectral radius (the largest absolute value of its eigenvalues). Then the infection will tend to increase when $R_0 > 1$ and to end if $R_0 < 1$.

\subsection{Physics-informed neural network}
\label{PINN}
Here, we use Physics-informed neural networks (PiNN) to solve the supervised learning problems, where physics laws presented as general nonlinear partial differential equations are given. By integrating the differential equations, as prior knowledge, into deep learning models, PiNN can bring us a data-driven forecasting, multi-physics model simulations and so on.

To be more specific on the construction of the PiNN, it generally takes fully connected layers, where each node in a
single layer is connected to every node in the following layer (except for the output layer), and each connection has a particular weight. A neuron takes the sum of weighted inputs from each incoming connection (plus a bias term), applies
an activation function, and passes the output to all the neurons in the next layer.
Mathematically, each neuron’s output looks as follows:
\begin{align*}
    g (\sum^n_{i=1} x_i w_i + b)
\end{align*}

where n represents the number of incoming connections, $x_i$ the value of each incoming neuron, $w_i$.
the weight on each connection, $b$ is a bias term, and $\sigma$ is referred to as the activation function.

Then, in addition to the data error, the residual of the differential equation is minimized in a least squares sense as part of the loss function.

\begin{figure}[H]
    \centering
    \includegraphics[scale=0.5]{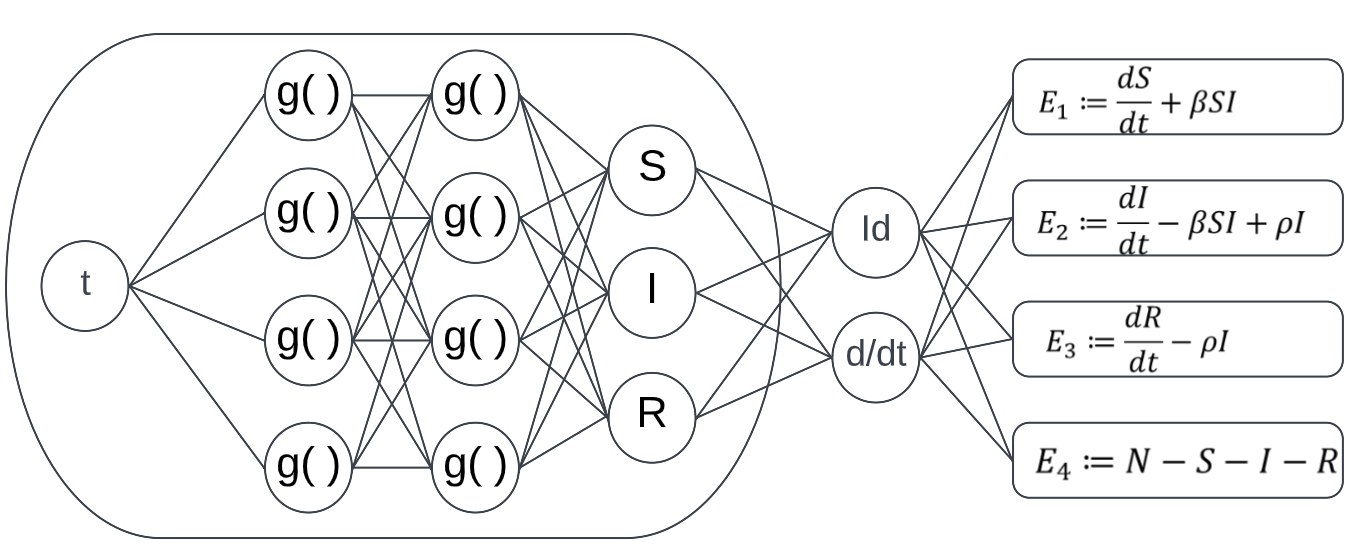}
    \caption[A two layers Physics-Informed Neural Networks]{A two-layer physics-informed neural network applied to the SIR (susceptible-infected-recovered) model, which takes time $t$ as an input and outputs the residuals E1 to E4. Using $g()$ to present the activation function.}
    \label{fig:pinn}
\end{figure}

Applying PiNN in the COVID-19 modeling, 
we acquire the derivatives to compute the residual networks E1, E2, E3 and E4 by applying the chain rule for differentiating compositions of functions using automatic differentiation.

In our formal computations, we employed a densely connected neural network, which takes the input variable t and outputs all components in the model (showing S, I, R in figure \ref{fig:pinn} for briefness). 

We employ automatic differentiation to obtain the required derivatives to compute the residual (physics-informed) networks E1, E2, E3 and E4. It is worth highlighting that parameters $\alpha$ and $\beta$ of the differential equations turn into parameters of the resulting physics informed neural networks E1, E2, E3 and E4. The total loss function is composed of the regression loss corresponding to I and the loss imposed by the differential equations system. For more information, refer to \cite{pinn-Raissi_Ramezani_Seshaiyer_2019}.

\begin{align}
\label{loss}
    Loss = MSE(S_{train}) +  MSE(I_{train}) + MSE(R_{train}) + \sum^4_{k=1} MSE(E_k)
\end{align}

\begin{align}
\label{error}
    Error = MSE(S_{test}) +  MSE(I_{test}) + MSE(R_{test}) + \sum^4_{k=1} MSE(E_k)
\end{align}

\subsection{Dataset}
Our datasets is organized from COVID-19 Data Repository by the Center for Systems Science and Engineering (CSSE) at Johns Hopkins University\cite{DONG2020533}, which is used widely by thousands of studies (Dong, E., 2020)\cite{DONG2020533}. After a branch of CSSE is forked, the following possible processes are implemented for data preprocessing:
(1)Data accumulation by country index;
(2)Removing negative outliers;
(3)Removing very high outliers(optional).

\section{Results}
\label{Implement}
\subsection{Parameter-driven simulation}

In this section, we are going to generate the data with prior-defined parameters sourced from papers (according to \cite{k_filter_math9060636} and \cite{ram2021modified}) and to check the hyper-parameters adjustment as well as computation complexity with our PiNN. (i.e. Applying PiNN to simulated data for different compartment models). The goal values for parameters refer to \cite{k_filter_math9060636} to make the simulation more like the infection situation in USA, which will be learned with the real-world data in the section \ref{data_dirven}. We choose 100 data from range (0,500) as our whole data size. General setting of hyper-parameters in the PiNN training, parameters learning results and loss are listed in the following tables.
    
\begin{table}[H]
    \caption{The PiNN hyperparameters: all models have the same data size, prediction ratio, learning rate scheduler, data type, and number of layers and nodes, but different training epochs (increasing with model complexity).}
    \label{tab:sim-set}
     \begin{center}
    \resizebox{1\linewidth}{\height}{
    \begin{tabular}{cccccccc}
    \hline
     Model & Data size & Prediction ratio & Learning rate scheduler & Data type & Number of layers (nodes) & Training epoch \\ \hline
     Basic SVEIDR & 100 & 0.9 & ReduceLROnPlateau \footnotemark  & Float32 & 8 (20) & 5.00E+5 \\
    Age-structured SVEIDR & 100 & 0.9 & ReduceLROnPlateau & Float32 & 8 (20) &  6.00E+5 \\
    Vaccination-structured SVEIDR & 100 & 0.9 & ReduceLROnPlateau & Float32 & 8 (20) & 8.00E+5 \\ \hline
    \end{tabular}
    }
    \end{center}
\end{table}
        
\footnotetext{A learning rate monitor that reduces learning rate when a metric has stopped improving}

\begin{table}[H]
    \caption{Parameters of SVEIDR models as described in section \ref{methods} and their goal, found, and accuracy values in PiNN.}
    \label{tab: parameter}
     \begin{center}
      \resizebox{1\linewidth}{1\height}{
    \begin{tabular}{ccccccc}
    \toprule
    Model & Parameter & Description & Goal Value (Training Range) &  Found Value & Accuracy(\%) \\ \midrule
    \multirow{7}{*}{Basic SVEIDR} & N (Stable) & The total number in an area & 3.5e7  & 3.5e7 &  \\
    & $\alpha$ & Vaccination rate & 3.5e-4 ($\pm$ 3.5e-3) & 0.00037 & 94.29\% \\
    & $\beta$ & Transmission rate & 0.3 ($\pm$ 0.5) & 0.3013 &  99.57\%\\
    & $\gamma$ & Incubation rate & 0.18 ($\pm$ 0.3) & 0.1808 &  99.56\%\\
    & $\sigma$ & Vaccine inefficacy & 0.05 ($\pm$ 0.1) & 0.0499 &  99.80\%\\
    & $\lambda$ &  Death rate & 0.06 ($\pm$ 0.1) & 0.0602 & 99.67\% \\
    & $\rho$ & Recovery rate & 0.1 ($\pm$ 0.5) &  0.1007 &  99.30\%\\
    \midrule
    \multirow{13}{*}{Aged SVEIDR} & N1 (Stable) & Total number of Age group 1 & $3.5e7
    \times 0.4$ &  $3.5e7
    \times 0.4$ &  \\
    & N2 (Stable) & Total number of Age group 2 & $3.5e7
    \times 0.6$ &  $3.5e7
    \times 0.6$ &  \\
    & $\alpha$ & Vaccination rate & 3.5e-4 ($\pm$ 3.5e-3) &  0.00029 &  82.86\%\\
    & $\beta$ & Transmission rate & 0.3 ($\pm$ 0.02) & 0.2959  & 98.63\%\\
    & $\gamma$ & Incubation rate & 0.18 ($\pm$ 0.02) & 0.1904 &  94.22\%\\
    & $\sigma$ & Vaccine inefficacy & 0.05 ($\pm$ 0.02) & 0.0569 &  86.20\%\\
    & $\lambda$ & Death rate & 0.06 ($\pm$ 0.02) & 0.0656  & 90.67\%\\
    & $\rho$ & Recovery rate & 0.1 ($\pm$ 0.02) & 0.1040 &  96.00\%\\
    & $M00$ & Age group 1's Contact coefficient & 20.7 ($\pm$ 0.1) &  20.6874 &  99.94\%\\ 
    & $M01$ &  Age group 1 and 2's Contact coefficient & 6.3 ($\pm$ 0.1) & 6.3012 &  99.99\%\\ 
    & $M11$ & Age group 2's Contact coefficient & 10.4 ($\pm$ 0.1) & 10.4127 & 99.87\%\\
     \midrule
     \multirow{15}{*}{Vaccinated SVEIDR} & N1 (Stable) & Total number of Age group 1 & $3.5e7
    \times 0.4$ & $3.5e7
    \times 0.4$  & \\
    & N2 (Stable) & Total number of Age group 2 & $3.5e7
    \times 0.6$ &  $3.5e7
    \times 0.6$ &  \\
    & $\alpha_1$ & Vaccination rate 1 & 3.5e-4 ($\pm$ 3.5e-3) &  0.00026  & 74.39\%\\
    & $\alpha_2$ & Vaccination rate 2 & 5e-4 ($\pm$ 5e-3) & 0.00039 &  78.00\%\\
    & $\beta$ & Transmission rate & 0.3 ($\pm$ 0.05) & 0.3039 &  98.7\%\\
    & $\gamma$ & Incubation rate & 0.18 ($\pm$ 0.02) & 0.1931 &  92.72\%\\
    & $\sigma_1$ & Vaccine inefficacy 1 & 0.05 ($\pm$ 0.02) & 0.0480  & 96.00\%\\
    & $\sigma_2$ & Vaccine inefficacy 2 & 0.15 ($\pm$ 0.02) & 0.1667 & 88.87\% \\
    & $\lambda$ & Death rate & 0.06 ($\pm$ 0.02) & 0.0536 &  89.33\%\\
    & $\rho$ & Recovery rate & 0.1 ($\pm$ 0.05) & 0.0943  & 94.30\%\\
    & $M00$ & Age group 1's Contact coefficient & 20.7 ($\pm$ 0.1) & 20.7214 &  99.89\%\\ 
    & $M01$ & Age group 1 and 2's Contact coefficient & 6.3 ($\pm$ 0.1) & 6.3199  & 99.68\%\\ 
    & $M11$ & Age group 2's Contact coefficient & 10.4 ($\pm$ 0.1) & 10.4650 & 99.38\%\\
    \bottomrule
    \end{tabular}
    }
    \end{center}
\end{table}
        
\begin{table}[H]
    \caption{Learning losses for SVEIDR models, including the total training loss, testing error, and the relative RMSE values for each component. Components S, D, and R exhibit best performance (RRMSE < 10\%).}
    \label{tab:loss}
    \begin{center}
    \resizebox{!}{0.9\height}{
    \begin{tabular}{ccccc}
    \toprule
    Model & Total training loss & Total error\footnotemark & Compartment &  Parameter-learned RRMSE(\%) \footnotemark \\ \midrule
    \multirow{6}{*}{Basic SVEIDR} &  \multirow{6}{*}{3.5543e-05} &
    \multirow{6}{*}{0.0003} & S & 0.04\% \\
    & & & V & 0.76\% \\
    & & & E & 0.24\% \\
    & & & I & 0.24\% \\
    & & & D & 0.16\%\\
    & & & R & 0.14\% \\ 
    \midrule
     \multirow{12}{*}{Aged SVEIDR} &
    \multirow{12}{*}{0.0005}& 
    \multirow{12}{*}{0.0031}
    & S1 & 0.17\% \\
    & & & V1 &  3.43\% \\
    & & & E1 & 0.86\% \\
    & & & I1 & 0.70\% \\
    & & & D1 & 0.31\%\\
    & & & R1 & 0.16\% \\ 
    & & & S2 & 0.14\% \\
    & & & V2 & 3.35\% \\
    & & & E2 & 0.75\% \\
    & & & I2 & 0.73\% \\
    & & & D2 & 0.32\%\\
    & & & R2 & 0.16\% \\ 
    \midrule
    \multirow{14}{*}{Vaccinated SVEIDR} &
    \multirow{14}{*}{0.0003} &
    \multirow{14}{*}{0.0051} 
    & S1 & 1.00\% \\
    & & & V11 &  2.61\% \\
    & & & V12 & 3.06\% \\
    & & & E1 & 3.43\% \\
    & & & I1 & 1.75\% \\
    & & & D1 & 0.32\% \\
    & & & R1 & 0.24\% \\ 
    & & & S2 & 0.83\% \\
    & & & V21 & 2.55\% \\
    & & & V22 & 3.59\% \\
    & & & E2 & 2.90\% \\
    & & & I2 & 1.83\% \\
    & & & D2 & 0.31\%\\
    & & & R2 & 0.25\% \\ 
    \bottomrule
    \end{tabular}
    }
    \end{center}
\end{table}

\footnotetext[2]{Total testing loss, specific function can refer Equation (\ref{loss})}

\footnotetext{Parameter learned Relative Root Mean Square Error, refer to Equation (\ref{relativemse})}

\subsubsection{Basic SVEIDR model}
\label{Imp: Basic}
The model construction for this section is shown in section \ref{Basic SVEIDR Model}. An 8-layer PiNN is introduced to training after we generate our simulated dataset with formula (\ref{f_basic}) and the goal values set in table \ref{tab: parameter}. 
The PiNN uses the information of the previous 0.9 ratio of data to calculate the loss function in training process, predict the last 0.1 ratio of data, and give the found values of parameters. Given that our models contain lots of parameters to be learned, limiting the training range for parameters, except for the stable parameter, is highly recommended. After that, we use the found parameters to generate a new disease data, also following formula (\ref{f_basic}), named as parameter-learned data, compare it with the original dataset, and calculate the Relative Root Mean Square Error (RRMSE) as Equation (\ref{relativemse})  for each compartment, which is a dimensionless form of Root Mean Square Error. According to \cite{Despotovic2016-pm} and \cite{McGough2017-oi}, model prediction accuracy is considered excellent if the RRMSE value is less than 10\%. Both the training loss and the RRMSE between parameter-learned and original data can be found in Table \ref{tab:loss}.

\begin{align}
\label{relativemse}
    RRMSE = \sqrt{\frac{\frac{1}{n} \sum_{i=1}^n (Y_i - \hat{Y_i})^2}{\sum_{i=1}^n Y_i^2}}
\end{align}

\begin{figure}[H]
    \centering
    \includegraphics[width=\linewidth]{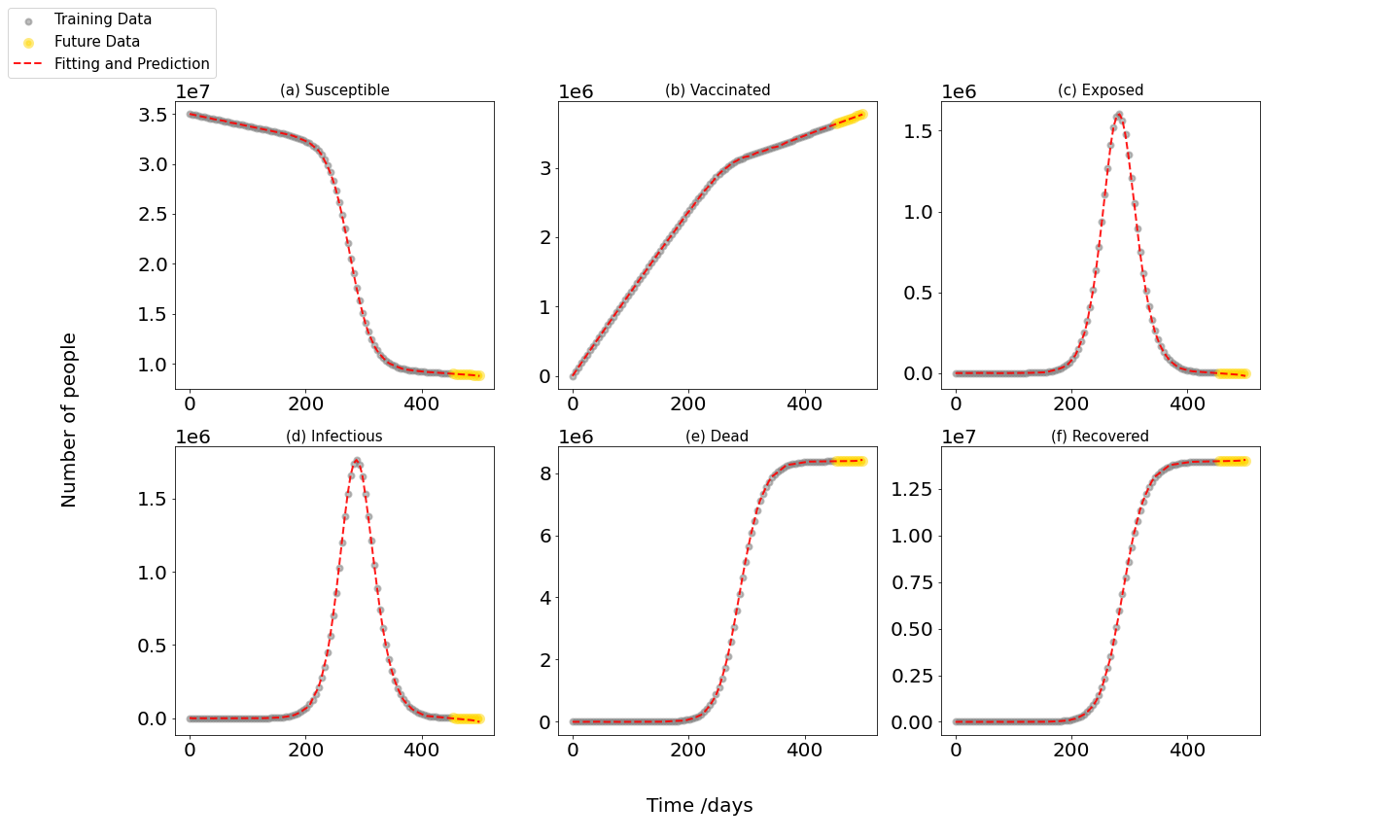} \caption{Basic SVEIDR model: prediction of the last 10\% of data for each component.} 
    \label{fig:Prediction plots (SVEIDR)}
\end{figure}

\begin{figure}[H]
    \centering
    \includegraphics[width=\linewidth]{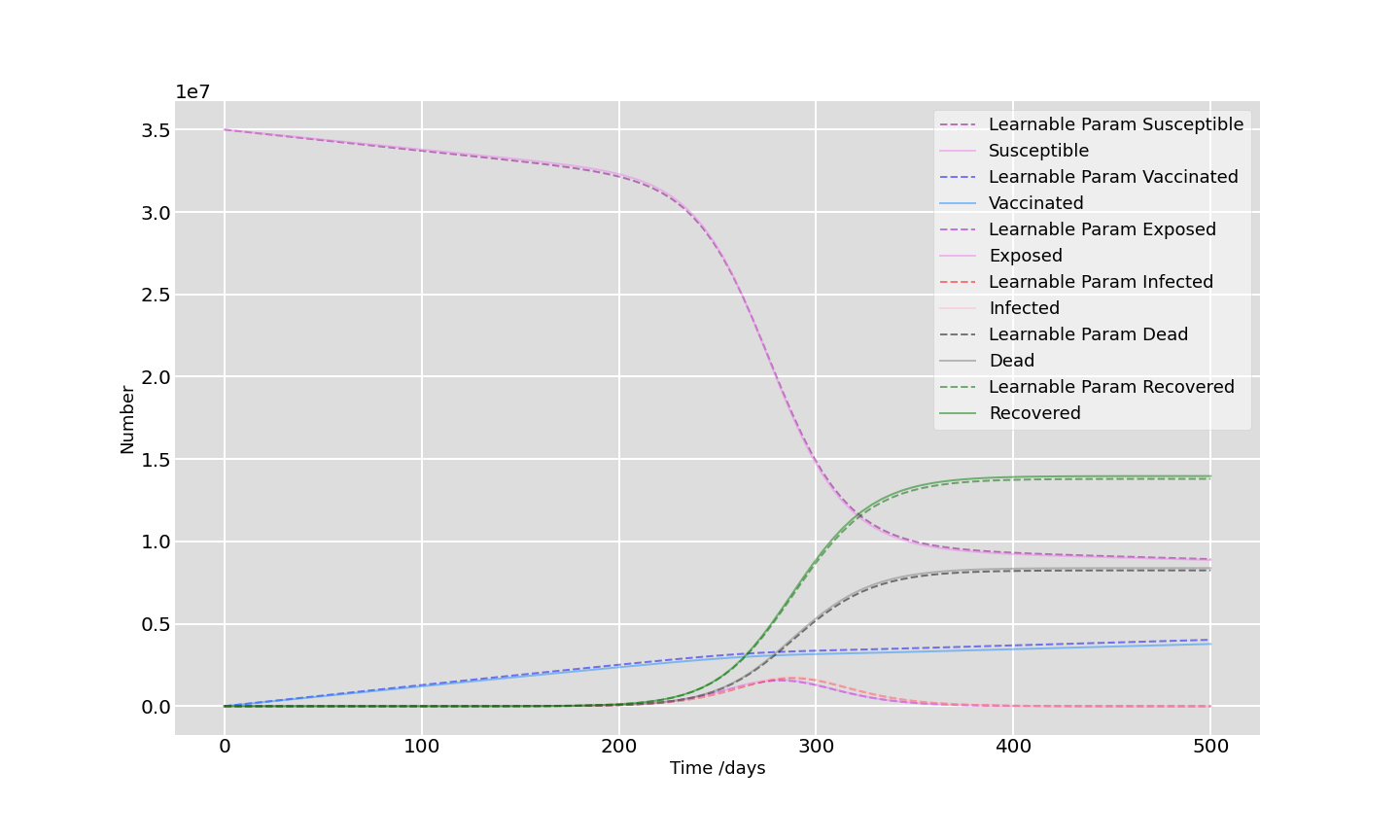} \caption{Basic SVEIDR model: comparison of the actual and found parameters for each component, with solid and dash lines representing the original and newly generated data, respectively.} 
    \label{fig:Parameter Learned Output (SVEIDR)}
\end{figure}

The prediction plots (Figure \ref{fig:Prediction plots (SVEIDR)}) show the forecasting results of each compartment (susceptible, vaccinated, exposed, infected, dead, and recovered), where the yellow color represents the last 0.1 ratio of data, i.e. future data, and the dash lines depict neural network's outcomes. The prediction dash lines fit the training data and predict the future data well. Further more, figure \ref{fig:Parameter Learned Output (SVEIDR)} shows the highly similarity of trends of parameter-learned and original data also tells the contribution of PiNN., where the solid and dash lines represent the original and newly generated data respectively. 

\subsubsection{Age-structured SVEIDR model}
\label{Imp: Age}
\begin{figure}[H]
    \centering 
    \noindent \makebox[\linewidth]{\includegraphics[width=\paperwidth]{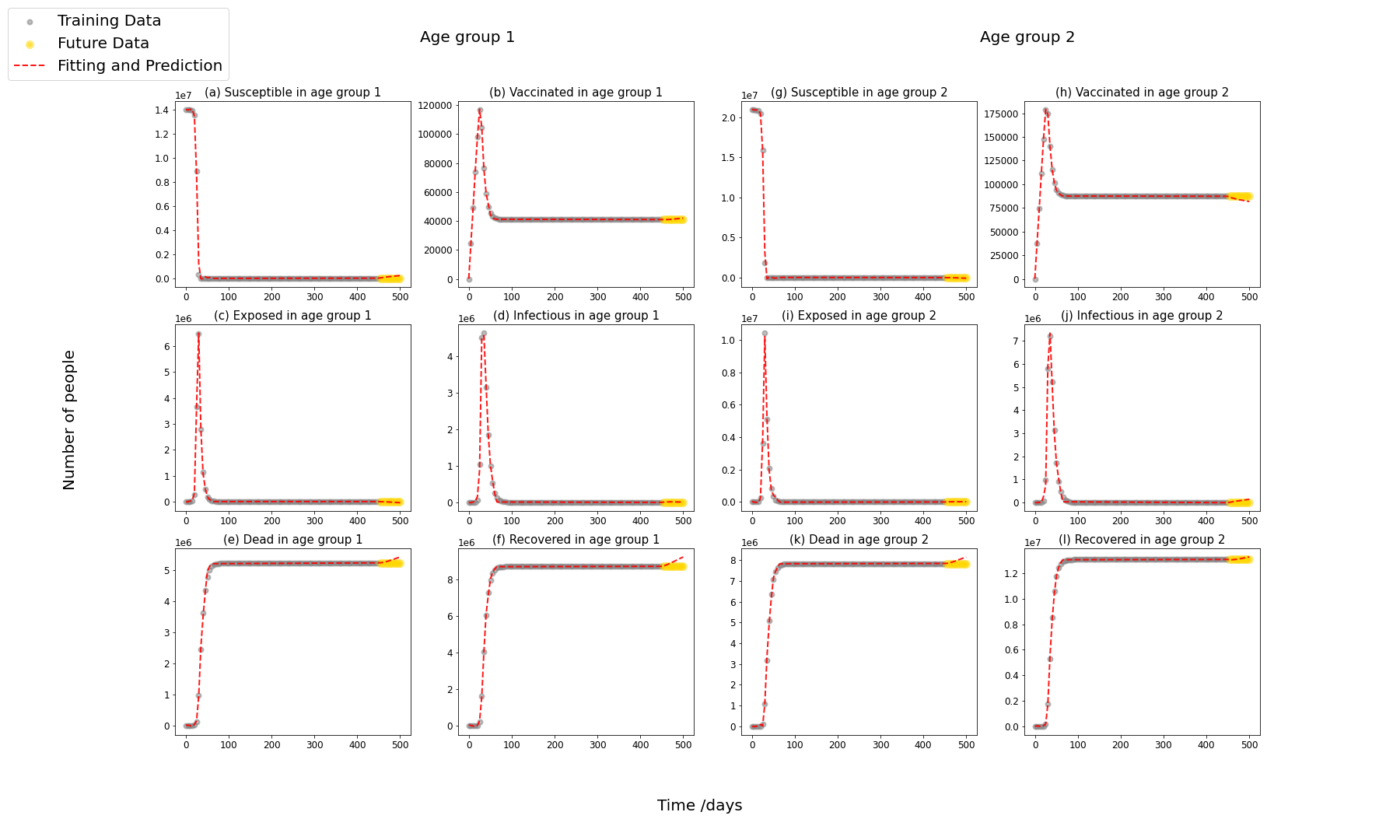}} \caption{Aged-structured SVEIDR model: prediction of the last 10\% of data for each component and age group.} \label{fig:Prediction plots (Aged)}
\end{figure}

In Age-structured SVEIDR model, people are divided. To avoid over complexity, we here simply choose 2 age groups, age 20-29 (age group 1) and 50-59 (age group 2), to form the contact matrix with their contact coefficients in \cite{ram2021modified}, and set the ratio of the number ($N1, N2$) of two groups to be 2:3. The experimental workflow is similar to the previous section \ref{Imp: Basic}. Figure \ref{fig:Prediction plots (Aged)} shows the PiNN's fit and forecast for all compartments in aged model, where the outcomes behave better on predicting S, V, E, I. Yet the parameter-learned RRMSE in table \ref{tab:loss} tells that found parameters perform better on S, D, R. For more information of the found parameters' output, refer to appendix \ref{appendix}.

\subsubsection{Vaccination-structured SVEIDR model}
\label{Imp: Vacc}
Based on the previous aged-structured SVEIDR model, we add 2 status (Doses) of Vaccination for this simulation, whose different equations can be found in section \ref{Vaccination structured Model}. Similarly, PiNN can predict well of S, V, E, I compartments, and tend to have better behavior on the E and I compartments in age group 1 than 2. The parameter learned output figures are included in appendix \ref{appendix}.

\begin{figure}[H]
    \centering  
    \noindent \makebox[\linewidth]{\includegraphics[width=1\paperwidth]{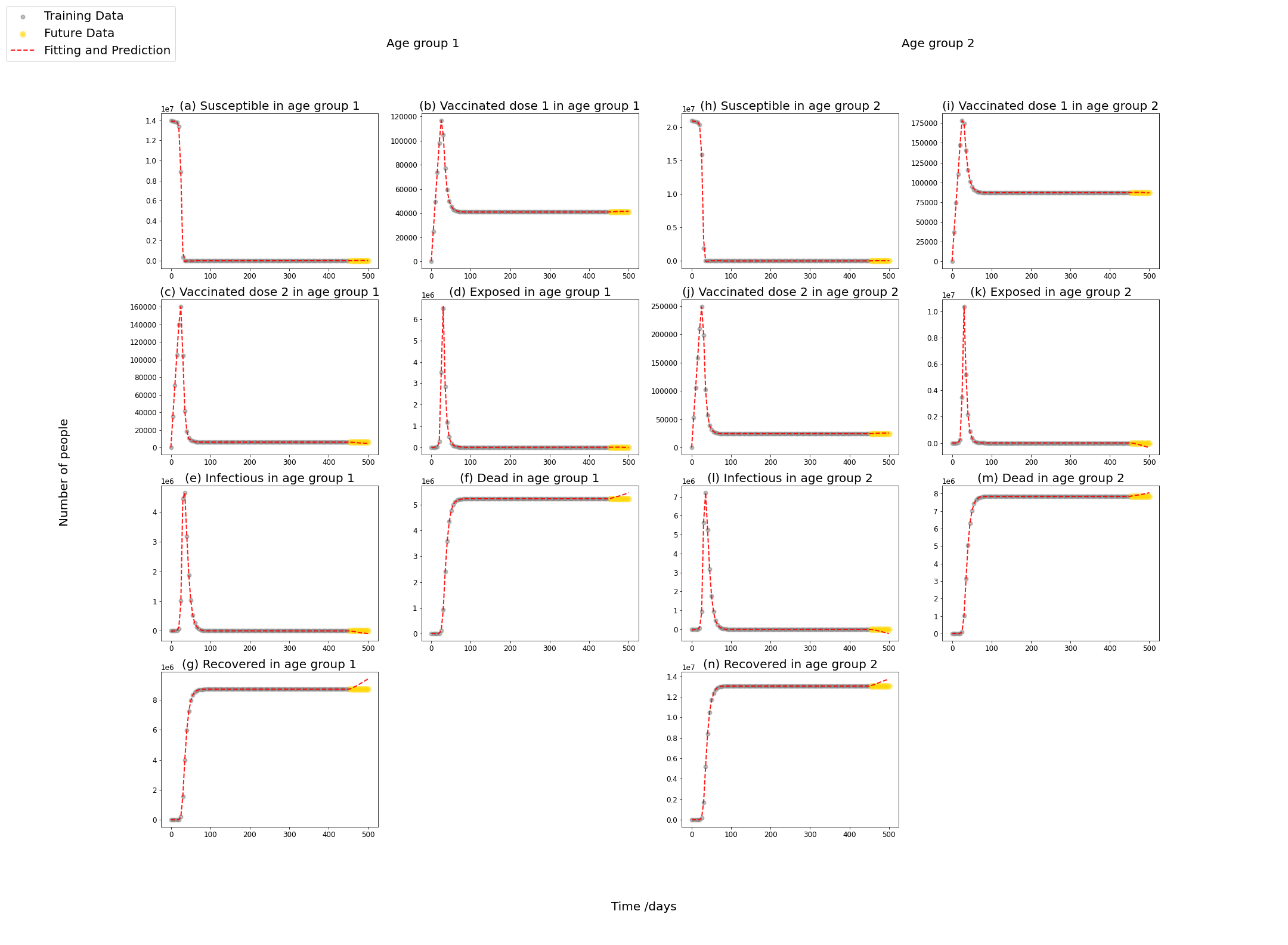} } \caption{Vaccination-structured SVEIDR model: prediction of the last 10\% of data for each component and age group. } \label{fig:Prediction plots (Vaccination)}
\end{figure}

\subsection{Data-driven experiment}
\label{data_dirven}
Different from the previous part, this part will focus on the modeling and prediction of the real data obtained in the data collation module.  Due to the limitation of real world data collection, we here present the epidemic situation with S, I, D, R compartments, where susceptible people directly transmit to the infected people. Then the transmission rate $\beta$ and the incubation rate $\gamma$ are combined into one coefficient, which we use incubation rate $\gamma$ to present in the following. The evaluation of the prediction will be carried out by testing errors.
    
\subsubsection{Basic information of dataset and its visualization}
After the time series visualization, we obtained an overview of the USA dataset time series as following Figure 8. It show us the dead active, recoverd, and infected population number in USA (Active cases = total cases - total recovered - total deaths. This value is for reference only after CSSE stopped to report the recovered cases). Comparing the data of the first 310 days with the data of the following days, the data of the previous days are more regular and unified, so we choose the first half for the experiment.
\begin{figure}[H]
    \centering
    \includegraphics[scale=1]{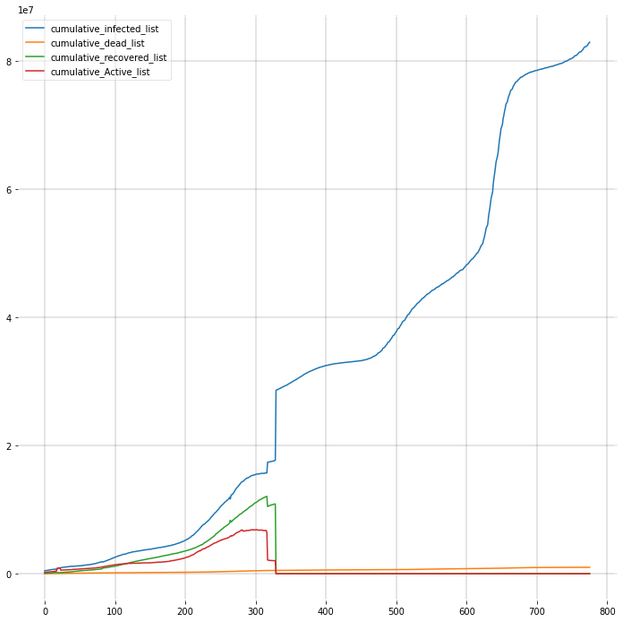}
    \caption{The US dataset time series, collected by Johns Hopkins University, for infected (blue), dead (orange), recovered (green), and active (red) people. Abrupt change is seen after day 310.}
    \label{fig:Overview of the US data set time series}
\end{figure}

Because only the initial case records follow the principle of recording the three types of data of illness, rehabilitation and death every day, We selected the first 310 days of data to model, timeserises of which is shown in the following Figure \ref{fig:Time series visualization of selected datasets}:
\begin{figure}[H]
    \centering
    \includegraphics[width=0.7\linewidth]{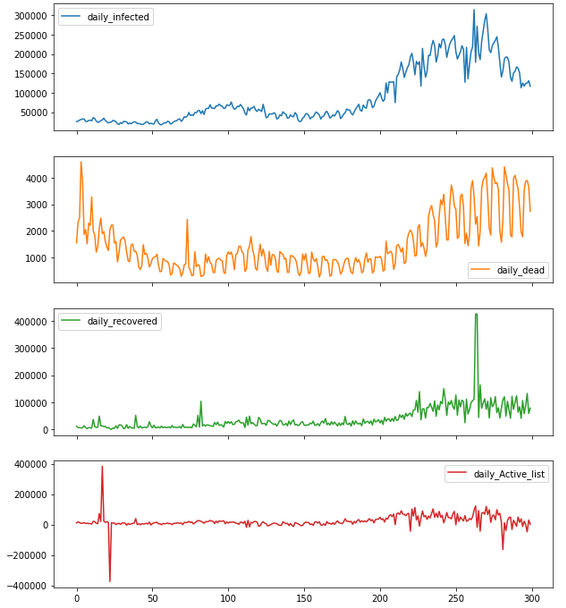}
    \caption{Time series visualization of selected datasets: showing the time series of daily infected, dead, recovered, active people in the first 310 days. }
\label{fig:Time series visualization of selected datasets}
\end{figure}
It can be seen from the figure that the real data show certain periodicity, outliers, noise and missing values, which show different characteristics from the data of parameter simulation. Therefore, we adopt different training strategies, which is partly shown in the Table 10.  And the whole record table of different trying is in the additional matrial.

\subsubsection{Result of USA data}
During training we optimize the process by conditioning the following training strategies and network settings:

\begin{table}[H]
\centering
\caption{Parameter setting and Learning results by PiNN on the USA data}
\label{tab:USA}
\begin{tabular}{|cc|ccc|}
\hline
\multicolumn{2}{|c|}{PiNN Parameter Setting}                 & \multicolumn{3}{c|}{Learning Results}                            \\ \hline
\multicolumn{1}{|c|}{Training size}                              & 290 (days)   & \multicolumn{2}{c|}{Training Loss}   & 1.68E-05 \\ \hline
\multicolumn{1}{|c|}{\multirow{2}{*}{\begin{tabular}[c]{@{}c@{}}Learning Rate\\ Scheduler\end{tabular}}} & \multirow{2}{*}{CyclicLR} & \multicolumn{2}{c|}{Total testing Error}  & 0.0087    \\ \cline{3-5} 
\multicolumn{1}{|c|}{}              &     & \multicolumn{1}{c|}{\multirow{3}{*}{Found parameters}} & \multicolumn{1}{c|}{Incubation rate} & $\gamma= 0.0130$   \\ \cline{1-2} \cline{4-5} 
\multicolumn{1}{|c|}{\multirow{2}{*}{Epochs}}    & \multirow{2}{*}{5E+05}    & \multicolumn{1}{c|}{}  & \multicolumn{1}{c|}{Death rate}      & $\lambda=0.0001$   \\ \cline{4-5} 
\multicolumn{1}{|c|}{}               &  & \multicolumn{1}{c|}{}      & \multicolumn{1}{c|}{Recovery rate}   & $\rho=0.0037$    \\ \hline
\multicolumn{1}{|c|}{\multirow{2}{*}{Layers}}  & \multirow{2}{*}{8}        & \multicolumn{1}{c|}{\multirow{4}{*}{Parameter-learned RRMSE(\%)}} & \multicolumn{1}{c|}{S}               & 0.01\% \\ \cline{4-5} 
\multicolumn{1}{|c|}{} & & \multicolumn{1}{c|}{} & \multicolumn{1}{c|}{I}               & 0.32\% \\ \cline{1-2} \cline{4-5} 
\multicolumn{1}{|c|}{\multirow{2}{*}{Nodes}} & \multirow{2}{*}{20}       & \multicolumn{1}{c|}{} & \multicolumn{1}{c|}{D} & 0.09\%       \\ \cline{4-5} 
\multicolumn{1}{|c|}{} & & \multicolumn{1}{c|}{} & \multicolumn{1}{c|}{R} & 0.27\%       \\ \hline
\end{tabular}
\end{table}

\begin{figure}[H]
    \centering \includegraphics[width=1\linewidth]{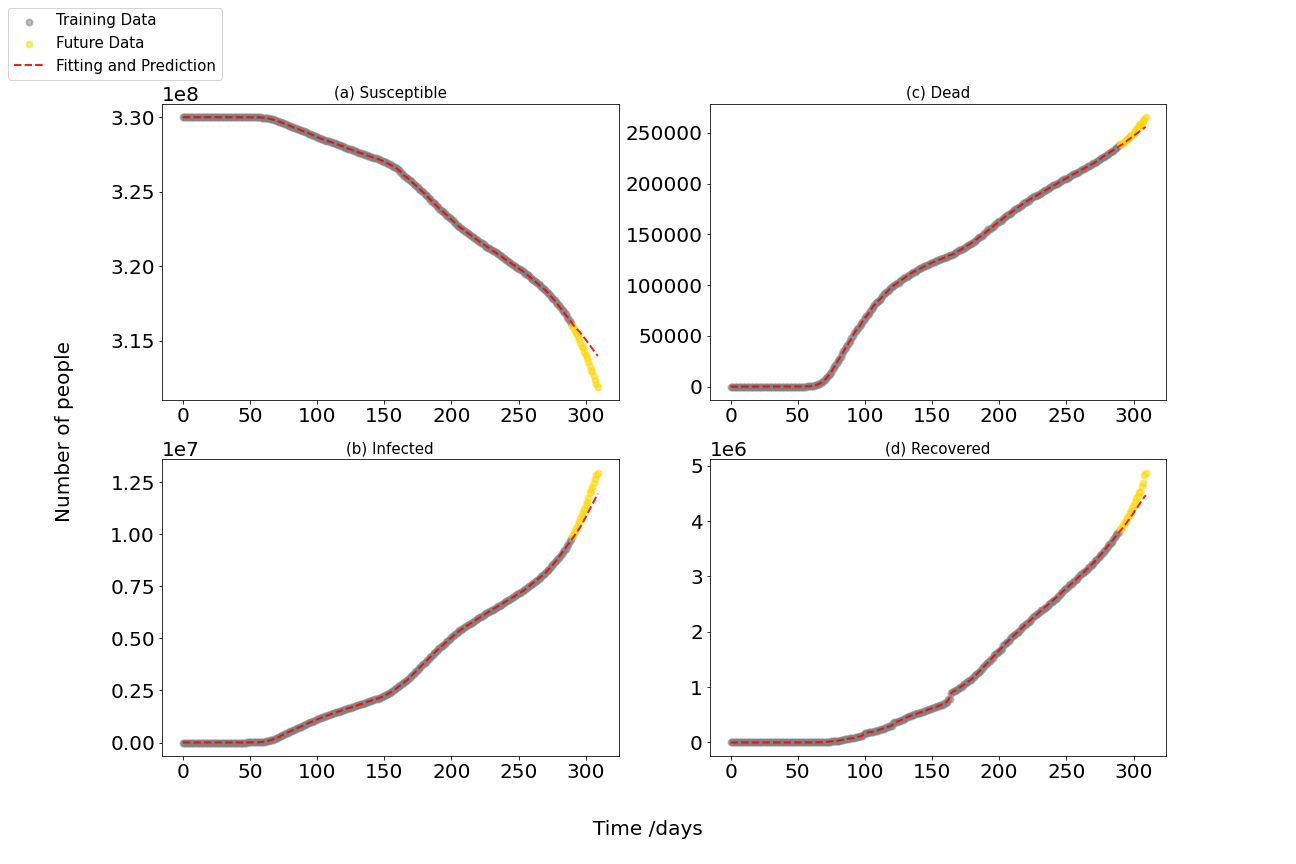}
    \caption{Model performance for the US dataset: prediction of the last 20 days of data for each component.}
    \label{Model fitting and prediction (USA)}
\end{figure}

Then the final modeling predictions in this section will be presented in the form of both hyper-parameters lists and prediction plots. The Figure \ref{Model fitting and prediction (USA)} is showing the forecasting results with the lowest training loss and prediction error in the previous table. 
From the prediction plots of USA data, we can find that the PiNN still fits the  training data fine and works roughly well on predicting the future trend of each components. The PiNN found incubation rate $\gamma= 0.0130$, death rate $\lambda=0.0001$, and recovery rate $\rho=0.0037$ with RRMSE ranging from 0.01\% to 0.35\% for the S, I, D, R compartments under this set.

\section{Discussion}
In the progress of implementation of PiNN on the simulated data, comparing the three models, we can tell that the loss and forecasting accuracy generally becomes worse when adding more components or parameters in the model (i.e. a more complex model), which can be simply described by the number of training epoch required to reach a roughly well prediction result. In this circumstance, one way to improve the accuracy without spending more time is to limit the range for the parameters learning, another is to adjust the learning rate scheduler.

Focusing on the parameter-learned RRMSE in table \ref{tab:loss}, we can find that the data generate by found parameters generally behave well on all components, with their RRMSE ranging from 0.01\% to 4\%. But all models tend to have much better performance on learning the S, D, R components rather than V, E, I, where RRMSE values are generally higher on V, E, I compartments. This can also be reflected in the found values of parameters, the vaccination rate $\alpha$, vaccine inefficacy  $\sigma$ and the incubation rate $\gamma$, related to V, E, I, tend to be farther from the goal values than others, which may be caused by having more complex interaction with other compartments in the model or the training range we set. However, prediction plots (figure \ref{fig:Prediction plots (SVEIDR)}, \ref{fig:Prediction plots (Aged)}, \ref{fig:Prediction plots (Vaccination)}) suggests another situation that PiNN is likely to have more testing error on D and R, which may due to the less informative of D and R for they link less with others in model.

\begin{figure}[H]
    \minipage{0.48\textwidth} \includegraphics[width=\linewidth]{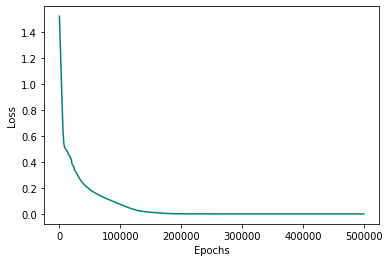} \subcaption{Training loss vs epoch with CyclicLR learning rate scheduler on the basic SVEIDR model} \label{fig:CyclicLR}
    \endminipage\hfill
    \minipage{0.48\textwidth}
     \includegraphics[width=\linewidth]{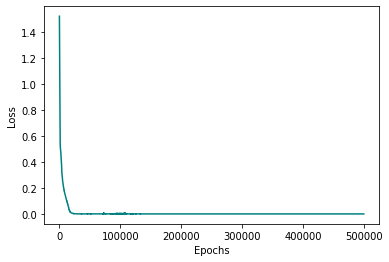}
      \subcaption{Training loss vs epoch with ReduceLROnPlateau learning rate scheduler on the basic SVEIDR model}
      \label{fig:ReduceLROnPlateau}
    \endminipage
    
    \minipage{0.48\textwidth} \includegraphics[width=\linewidth]{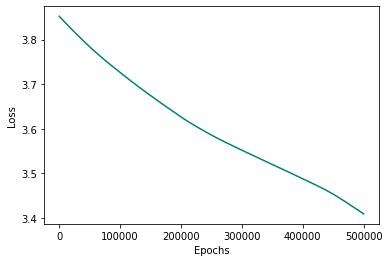} \subcaption{Training loss vs epoch with CyclicLR learning rate scheduler on the aged SVEIDR model} \label{fig:CyclicLR2}
    \endminipage\hfill
    \minipage{0.48\textwidth}
     \includegraphics[width=\linewidth]{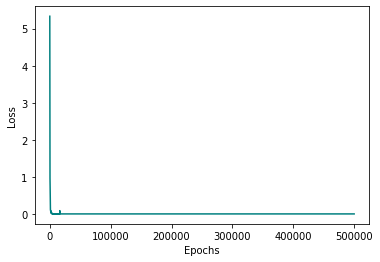}
      \subcaption{Training loss vs epoch with ReduceLROnPlateau learning rate scheduler on the aged SVEIDR model}
      \label{fig:ReduceLROnPlateau2}
    \endminipage
    \caption{Training loss vs epoch for comparison of CyclicLR and ReduceLROnPlateau scheduler using in the basic and aged SVEIDR model.}
\end{figure}

Taking learning rate scheduler into account, this paper compare two types of monitor: CyclicLR and ReduceLROnPlateau on the basic and age-structured SVEIDR model. CyclicLR scheduler makes learning rate cycle between a set boundaries with a certain frequency, while ReduceLROnPlateau scheduler  monitors a quantity and decays the learning rate when the quantity stops improving. Figure \ref{fig:CyclicLR} and Figure \ref{fig:ReduceLROnPlateau} are the training loss plots comparison for the two method in \ref{Imp: Basic}. Figure \ref{fig:CyclicLR2} and Figure \ref{fig:ReduceLROnPlateau2} are the training loss plots comparison for the two method in \ref{Imp: Age}. Within $5.00E+5$ training epochs, ReduceLROnPlateau scheduler shows obviously higher converging speed on both two models while CyclicLR will have a more smooth loss decay.

Unlike traditional mathematical computing such as the Euler and Longkutta methods and the Long-Short Term Memory that purely rely on machine learning and deep learning, the PiNN we used have certain advantages, including wider practicability, more software and hardware support, and more convenient model structural modification.

\section{Conclusions}
In this paper, we present an enhanced predictive modeling for COVID-19 pandemic. We improve the existing compartmental model to incorporate major driving parameters such as vaccination, social contact among different age groups. We provide detailed model analysis as we derive the basic SVEIDR, age group SVEIDR to prove non-negativity and determine basic reproduction number. We learn the model parameters using a data driven approach that incorporates real life epidemiological scenario. We also predict real data to better demonstrate the availability and performance of this disease informed neural network. The predictive errors (RRMSE)  can be found significantly small to be less than 4\% on parameter-driven simulation with the proposed models.
Also, PiNN finds incubation rate $\gamma= 0.0130$, death rate $\lambda=0.0001$, and recovery rate $\rho=0.0037$ for the first 310 days of epidemic in USA with less than 0.35\% Relative RMSE error for all compartments.
In terms of the training time, basic, aged, vaccinated models take about 40mins, 100mins, 210mins respectively, where the ReduceLROnPlateau learning rate schedule plays an essential role for the converging speed. Further, vaccinated component shows difficulty more than others while training. While the current model is demonstrated for COVID-19, it can be extended to model other pandemic scenarios in the future for timely intervention and prevent global loss.

\section*{Acknowledgements}
\noindent This work was supported by Touch Education Technology Inc. We acknowledge scientific support from Dr. R. Singh of Massachusetts Institute of Technology; editorial support from Dr. J. S. Lim of Harvard University; and administrative support from C. Ding of Touch Education Technology Inc.

\section*{Author contributions}
\noindent J.K. and J.M. performed modeling of small-scale dataset. J.K. conducted mathematical derivations and implemented parameter-driven simulation part. J.M. performed real-world data collection, processing, visualization, and prediction. X.Y. focused on literature reviews. R.S worked on conclusion. All authors contributed to manuscript preparation.

\section*{Competing financial interests}
\noindent The authors declare no competing financial interests.

\bibliography{reference}
\bibliographystyle{abbrv}

\appendix
\section{Appendix figures}
\label{appendix}

\begin{figure}[H]  \includegraphics[width=\linewidth]{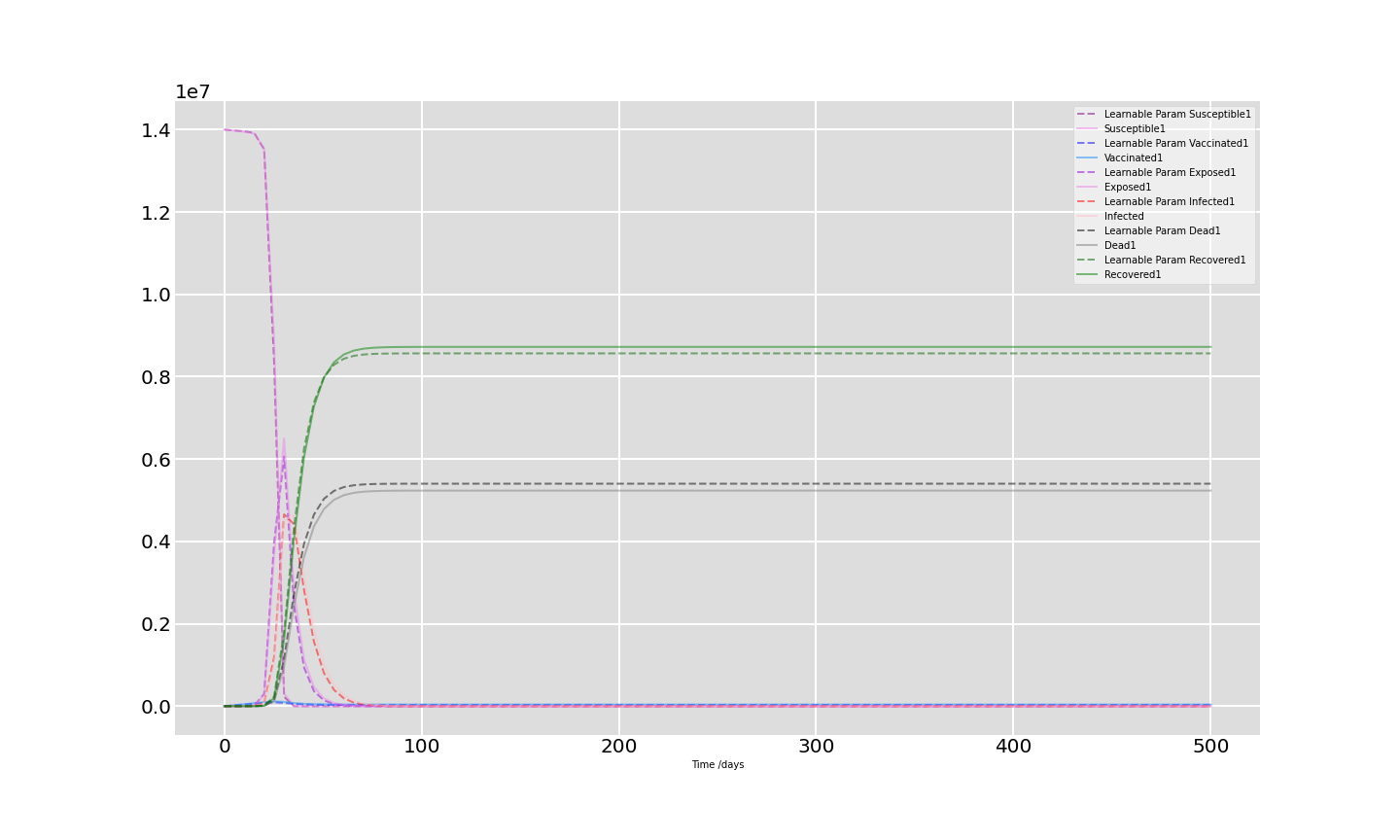} \caption{Parameter Learned Output (Age group 1)} \label{fig:Parameter Learned Output (Age group 1)}
\end{figure}

\begin{figure}[H]
 \includegraphics[width=\linewidth]{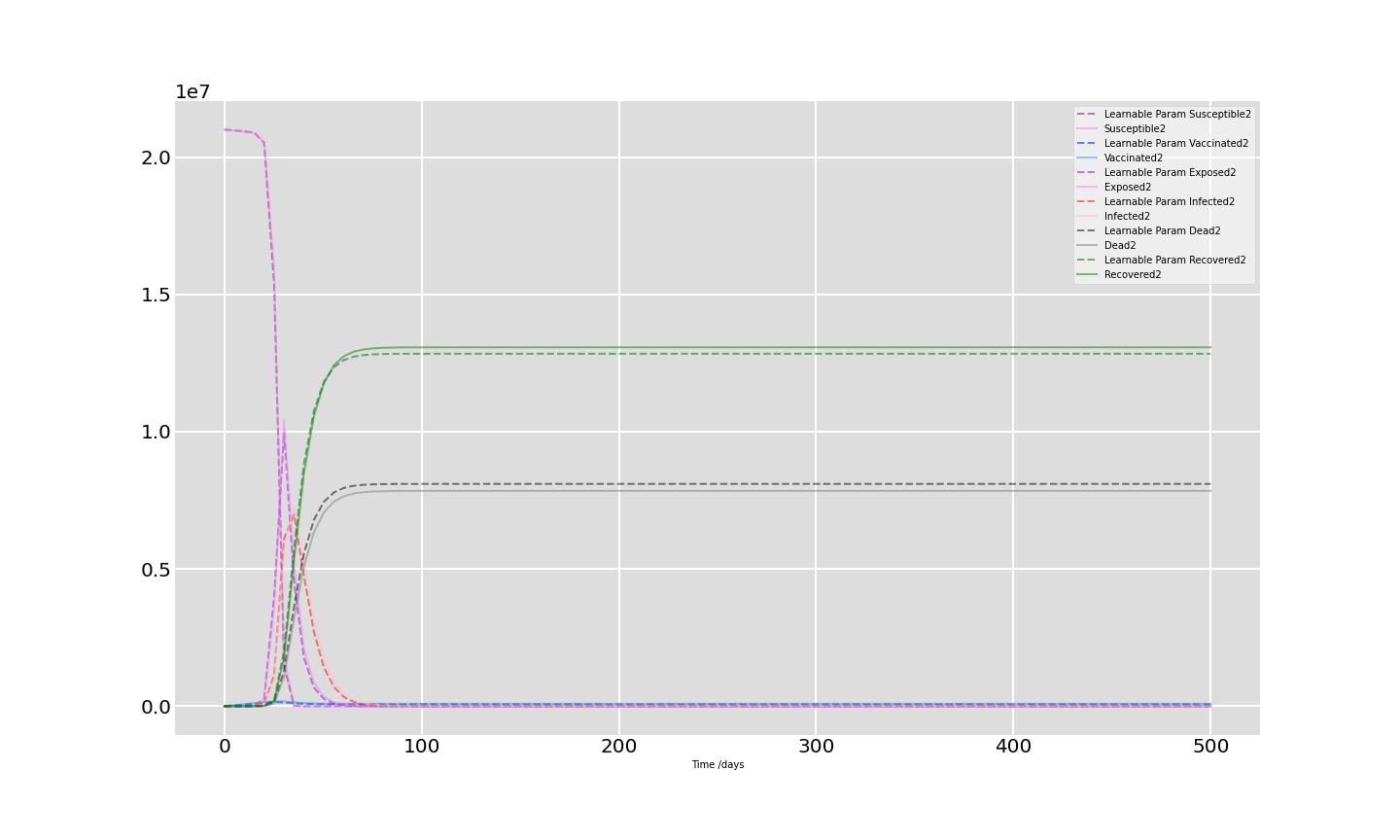} \caption{Parameter Learned Output (Age group 2)} \label{fig:Parameter Learned Output (Age group 2)}
\end{figure}
     
\begin{figure}[H]  \includegraphics[width=\linewidth]{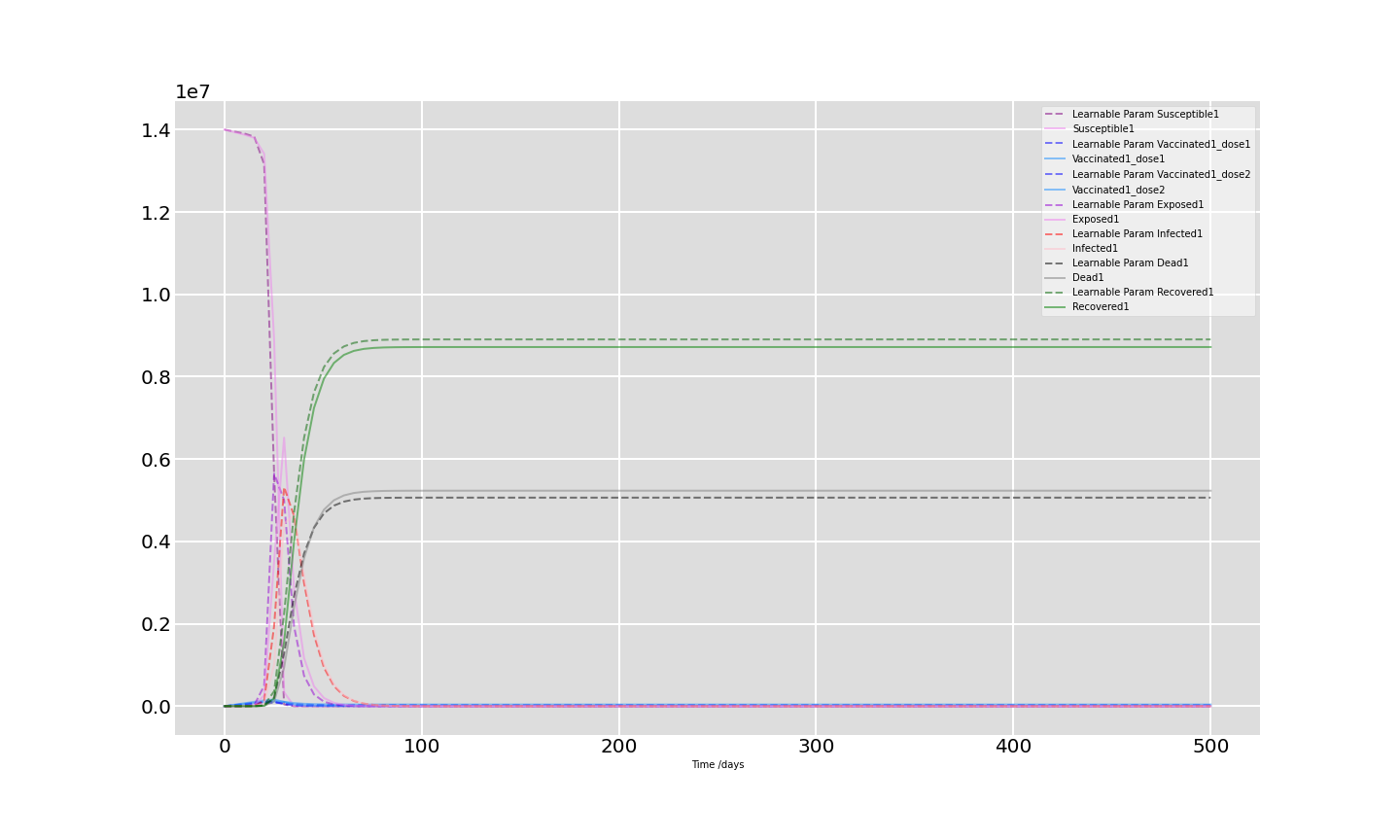} \caption{Parameter Learned Output (Vaccination-structured aged group 1)} \label{fig:Parameter Learned Output (Vaccination group 1)}
\end{figure}

\begin{figure}[H]
 \includegraphics[width=\linewidth]{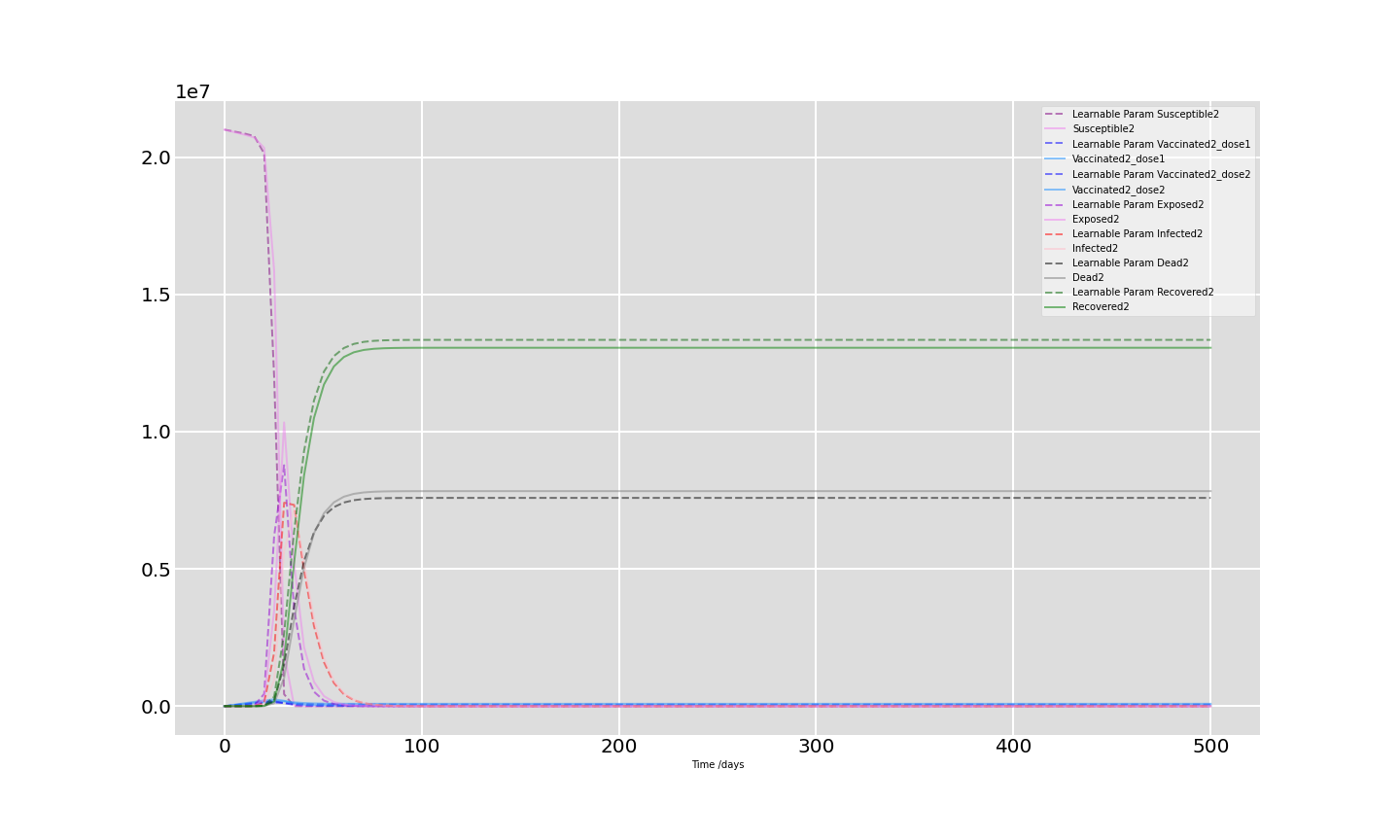}
 \caption{Parameter Learned Output (Vaccination-structured aged group 2)}
\label{fig:Parameter Learned Output (Vaccination group 2)}
\end{figure}
        
\end{document}